\documentclass[prb,twocolumn,showpacs,preprintnumbers,amsmath,amssymb,floats]{revtex4-1}
\usepackage{epsfig}
\usepackage{setspace}
\usepackage{graphicx}
\usepackage{amsmath}
\usepackage{epstopdf}
\usepackage{amsbsy}
\usepackage{dcolumn}
\usepackage{bm}

\usepackage[usenames]{color}

\begin{document}

\title{Low-energy bound states at interfaces between superconducting and block antiferromagnet regions in K$_x$Fe$_{2-y}$Se$_2$}
\author{S. Mukherjee$^{1,2}$, M. N. Gastiasoro$^1$, P. J. Hirschfeld$^3$, B. M. Andersen$^1$}
\affiliation{
$^1$Niels Bohr Institute, University of Copenhagen, DK-2100 Copenhagen \O, Denmark\\
$^2$Niels Bohr International Academy, Niels Bohr Institute, University of Copenhagen, Blegdamsvej 17, DK-2100 Copenhagen \O, Denmark\\
$^3$Department of Physics, University of Florida, Gainesville, Florida 32611, USA}


\begin{abstract}

The high-T$_c$ alkali doped iron selenide superconductors K$_x$Fe$_{2-y}$Se$_2$ have been recently shown to be intrinsically phase separated into Fe vacancy ordered block antiferromagnetic regions and superconducting regions at low temperatures. In this work, we use a microscopic five orbital Hubbard model to obtain the electronic low-energy states near the interfaces between block antiferromagnets and superconductors. It is found that abundant low-energy in-gap bound states exist near such interfaces irrespective of whether the superconductor has $d$- or $s$-wave pairing symmetry. By contrast, it is shown how nonmagnetic scattering planes can provide a natural means to distinguish between these two leading pairing instabilities of the K$_x$Fe$_{2-y}$Se$_2$ materials.

\end{abstract}

\pacs{74.70.Xa, 74.20.Rp, 74.55.+v, 74.81.-g}

\maketitle
\section{Introduction}
Alkali doped iron selenide materials\cite{Hirschfeld:2011, Dagotto:2012} A$_x$Fe$_{2-y}$Se$_2$ (A=alkali element) are high temperature superconductors with T$_c\sim32$K\cite{Guo:2010} that also undergo a transition to a iron vacancy ordered structure at T$_S\sim 578K$\cite{Zavalij:2011,Bao:2011,Pomjakushin:2011,Ye:2011} followed closely by a transition to a block anti-ferromagnetic state (BAFM) with Neel temperatures T$_N\sim 559K$.\cite{Bao:2011,Pomjakushin:2011,Ye:2011} The  BAFM develops within the $\sqrt{5}\times\sqrt{5}$ arrangement of iron vacancies with a magnetic moment of $m\sim3.31\mu_B/Fe$. This is the highest reported magnetic moment among all iron pnictide and iron chalcogenide superconductors. The existence of the above phases within the superconducting state has generated intense debate about the interplay between these seemingly competing phases, and about the origin of superconducting pairing leading to high temperature superconductivity in this unusual environment.\cite{Mazin:2011}

The presence of phase separation between superconducting regions and the BAFM has been confirmed using a variety of experiments including spectroscopy,\cite{Chen:2011,Lazarevic:2012,Yuan:2012} microscopy,\cite{Chen:2011,Yuan:2012,Ding:2012,Li(a):2012,Li(b):2012,Yan:2012,Landsgesell:2012,Charnukha:2012} and x-ray diffraction\cite{Landsgesell:2012,Ricci:2011,Liu:2012} techniques.
The phase separation has been observed along the $c$-axis and also in the FeSe plane as nanoscale filamentary superconducting domains existing within the vacancy ordered BAFM regions.\cite{Ding:2012,Ricci:2011,Yuan:2012} Though the morphology and physical properties of the metallic nano domains can depend on the thermal history\cite{Landsgesell:2012,Liu:2012} and stoichiometry,\cite{Yan:2012} a number of alkali iron selenide samples have been observed to be free from iron vacancies in the superconducting regions.\cite{Lazarevic:2012,Yan:2012,Texier:2012}

Angular resolved photoemission (ARPES) experiments investigating the metallic regions observe a Fermi surface with two electron pockets around the M points and a small electron pocket developing around the Z point.\cite{Wang:2011, Zhang:2011,Qian:2011} The absence of a hole pocket around the $\Gamma$ point has reinvigorated the discussion of the origin of superconductivity in the Fe-based materials. Specifically, the absence of $(\pi,0)$ nesting between hole and electron Fermi surface sheets prevents the leading $s\pm$ pairing instability found e.g. within spin fluctuation mediated pairing.\cite{Mazin:2008} Theoretical calculations for the superconducting ground state are currently inconclusive and predictions have been made for both $d$-wave \cite{Maier:2011, WangF:2011} and $s$-wave \cite{Fang:2011,Mazinb:2011} pairing symmetry.

Measurements within the superconducting state find a nodeless superconducting gap.\cite{Wang:2011,Zhang:2011,Li(b):2012,Zeng:2011} Although ARPES detects a nearly isotropic gap structure,\cite{Wang:2011} STM measurements have observed a double gap feature in the local density of states (LDOS).\cite{Li(a):2012,Li(b):2012} This does not identify the gap  symmetry in alkali iron selenides since neither $d$-wave nor $s$-wave symmetry possess any symmetry enforced nodes at $k_z=0$ for this Fermi surface. However, ARPES measurements in Ref.~\onlinecite{Xu:2012} find an isotropic superconducting gap structure at the Z point where a electron pocket exists around the $\Gamma$ point indicating an $s$-wave gap symmetry.

From the above discussing, it is evident that new theoretical methods to determine the pairing symmetry of these materials are desirable. In this paper we study the in-plane interface states near boundary regions of a BAFM with ordered arrangement of iron vacancies and a next nearest neighbor (nnn) $s$-wave or nearest neighbor (nn) paired $d$-wave superconductor. Note that in the absence of a Fermi pocket around the $\Gamma$ point the nnn $s\pm$ state does not have a sign change between Fermi pockets. We denote this superconducting state by $s$-wave with the understanding that it represents nnn superconducting pairing interaction in real space. We use a five orbital tight binding Hamiltonian and include superconductivity by generating pairing from a spin fluctuation exchange mechanism within an RPA weak coupling theory. The model is able to reproduce the electronic structure of K$_x$Fe$_{2-y}$Se$_2$ and stabilise a BAFM structure as well as a superconducting state with $s$-wave or $d$-wave symmetry. It is found that the presence of an interface between these two ordered phases leads to the formation of in-gap bound states which should be detectable by future STM measurements. Bound interface states are generated for both $s$-wave and $d$-wave symmetry due to the magnetic nature of the BAFM. A qualitative difference between $s$- and $d$-wave order is shown to arise near e.g. (110) non-magnetic scattering boundaries such as cracks, grain boundaries or free surfaces within the superconducting phase.

The paper is organized as follows, in the next section we provide details of the model used for our calculations. This is followed by an analysis of the
electronic structure, magnetic order, bulk superconducting order and finally we discuss the results for the LDOS in the simplified inhomogeneous situation of an interface between a BAFM and a superconductor.

\section{Model}

\begin{figure}[]
\begin{minipage}{.99\columnwidth}
\includegraphics[clip=true,width=0.99\columnwidth]{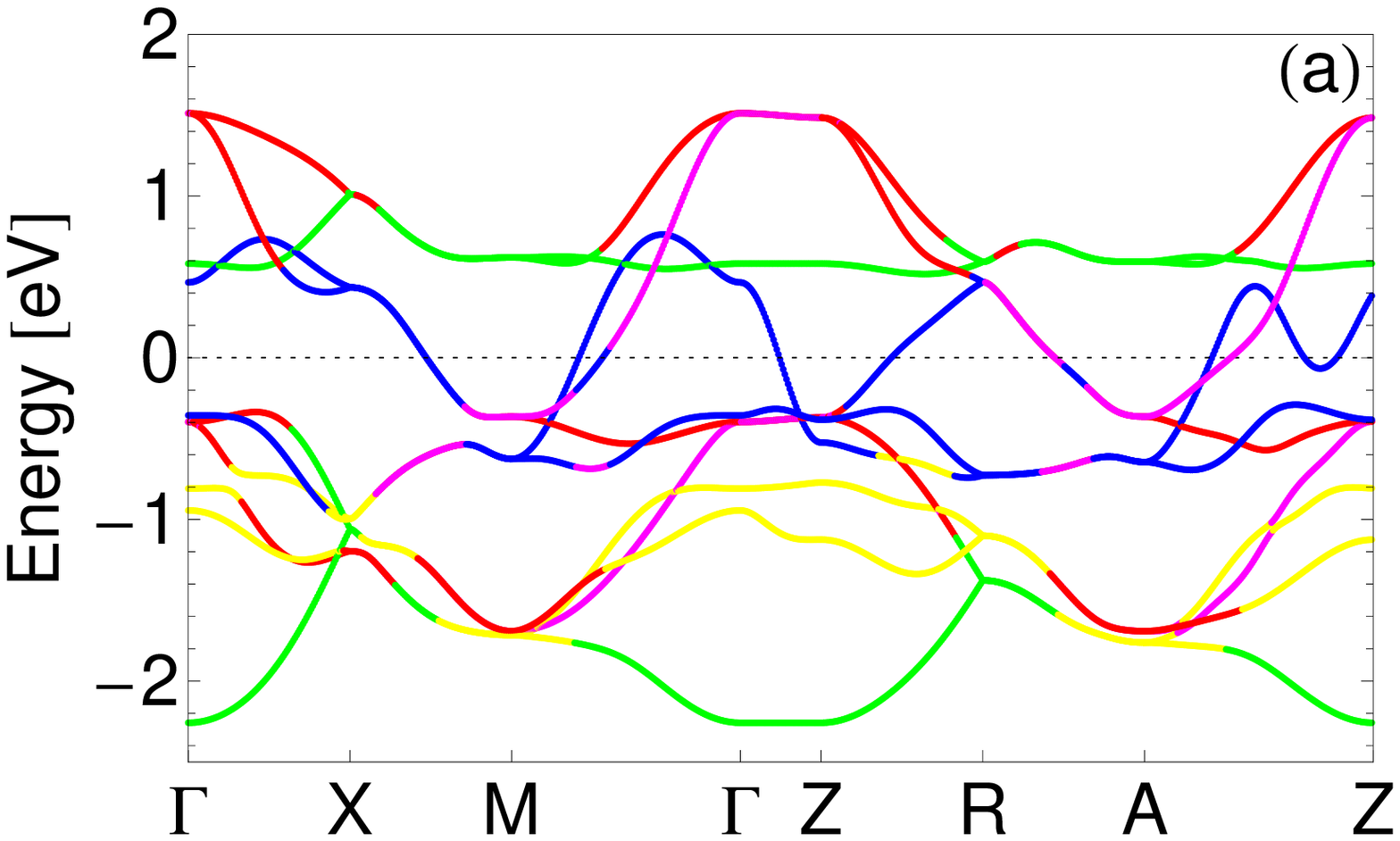}
\end{minipage}
\begin{minipage}{.49\columnwidth}
\includegraphics[clip=true,width=0.99\columnwidth]{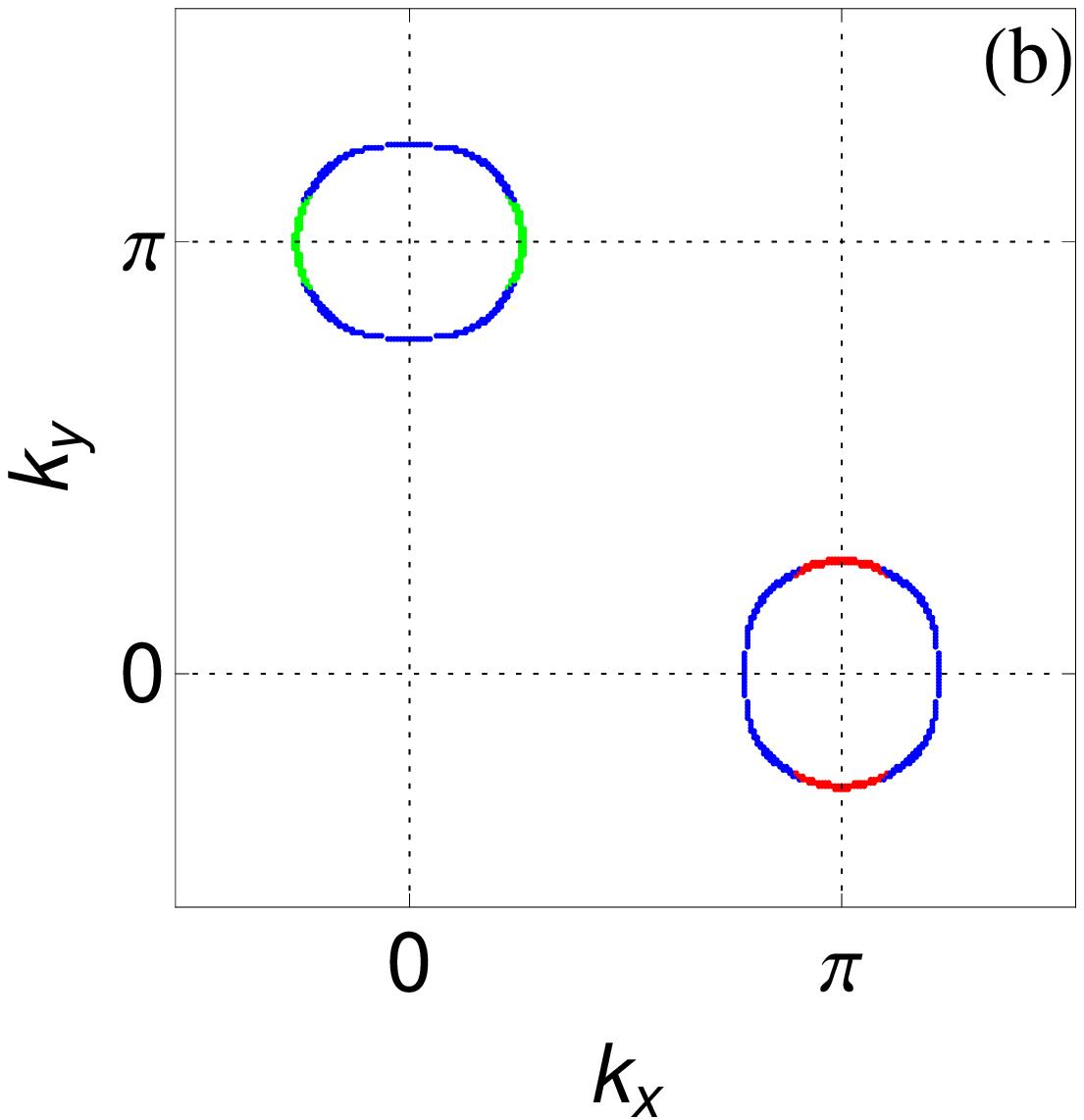}
\end{minipage}
\begin{minipage}{.49\columnwidth}
\includegraphics[clip=true,width=0.99\columnwidth]{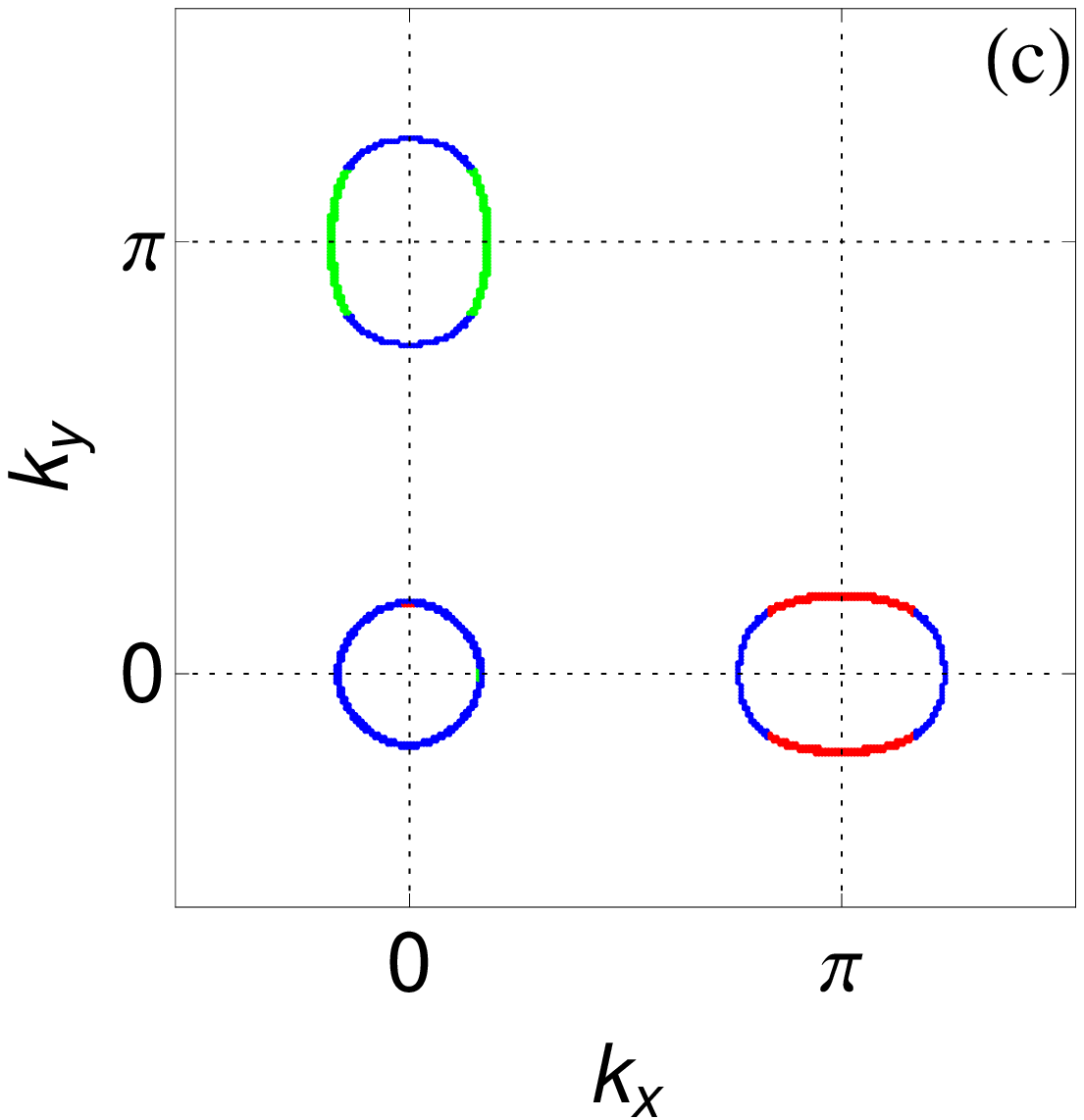}
\end{minipage}
\begin{minipage}{.99\columnwidth}
\includegraphics[clip=true,width=0.99\columnwidth]{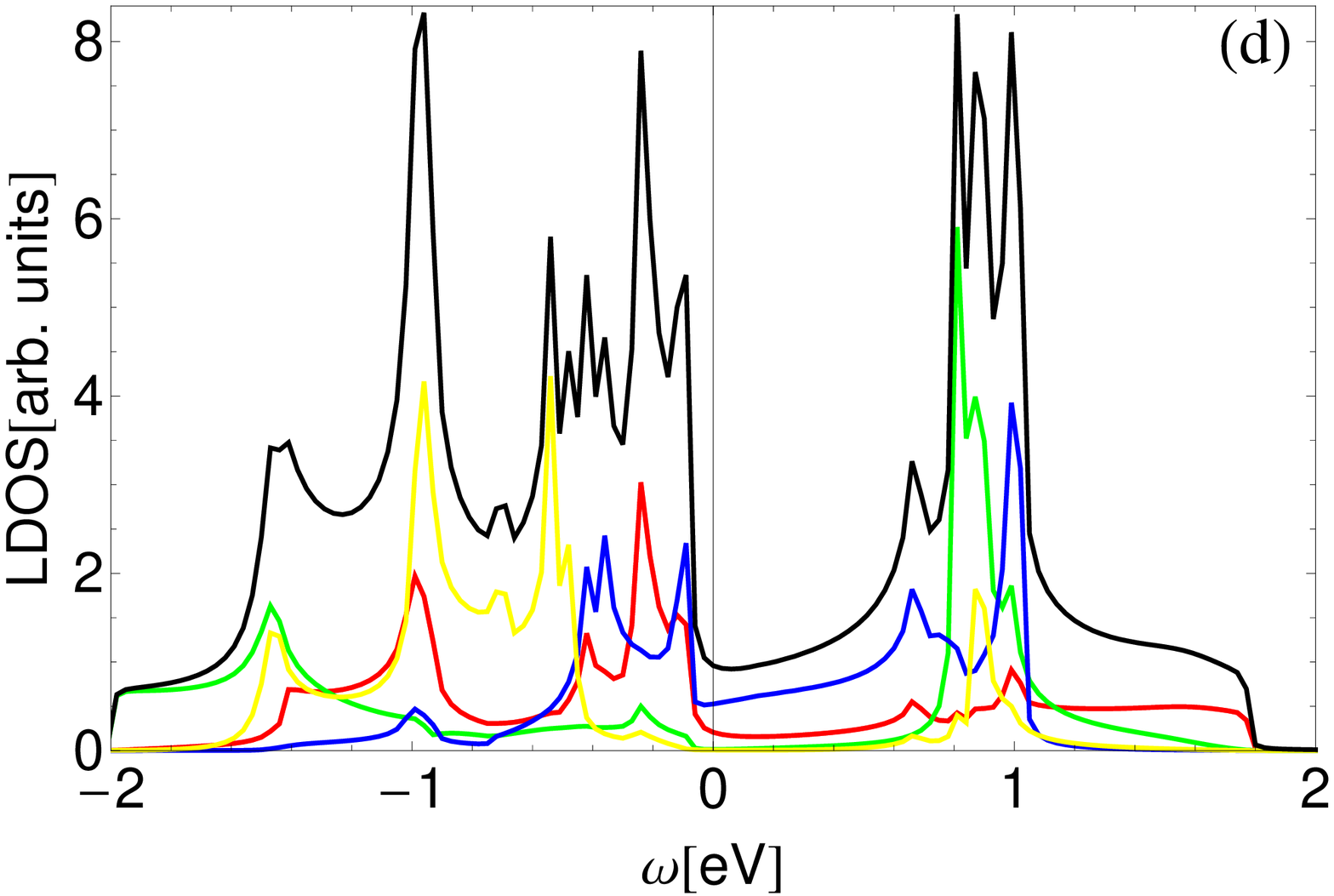}
\end{minipage}
\caption{(Color online) (a) Band structure of K$_x$Fe$_{2-y}$Se$_2$. Here the color codes represent green: $d_{xz}$, red: $d_{yz}$, cyan: $d_{x^2-y^2}$, blue: $d_{xy}$, yellow: $d_{z^2}$.
(b) Fermi surface calculated at $k_z=0$. (c) Fermi surface at $k_z=\pi$. (d) Orbital resolved LDOS at $k_z=0$ in the normal state at an electron doping of $n=0.15$ e$^{-}$/Fe.} \label{fig:normal}
\end{figure}

The five-orbital model Hamiltonian is given by
\begin{equation}
 \label{eq:H}
 H=H_{0}+H_{int}+H_{SC}+H_{vacancy}.
\end{equation}
The first term is determined by the tight-binding band
\begin{equation}
 \label{eq:H0}
H_{0}=\sum_{\mathbf{ij},\mu\nu,\sigma}t_{\mathbf{ij}}^{\mu\nu}c_{\mathbf{i}\mu\sigma}^{\dagger}c_{\mathbf{j}\nu\sigma}-\mu_{0}\sum_{\mathbf{i}\mu\sigma}n_{\mathbf{i}\mu\sigma}.
\end{equation}
Here the operators $c_{\mathbf{i} \mu\sigma}^{\dagger}$ ($c_{\mathbf{i}\mu\sigma}$) create (annihilate) an electron
at the $i$-th site in the orbital $\mu$ and with spin projection $\sigma$, and $\mu_{0}$ is the chemical potential.
The indices $\mu$ and $\nu$ correspond to the $d_{xz}$, $d_{yz}$, $d_{x^2-y^2}$, $d_{xy}$
and $d_{3z^2}$ iron orbitals.
The hopping parameters $t_{\mathbf{ij}}^{\mu\nu}$ are obtained by slightly modifying the tight-binding parameters of KFe$_2$Se$_2$
in Ref.~\onlinecite{Yong:2012}. This is similar to the prescription used by Maier {\it et al.}\cite{Maier:2011} and consists of
shifting the $d_{xz}$/$d_{yz}$ orbital nearest neighbor hopping by -0.08 eV and $d_{xy}$ orbital hoppings by 0.08 eV.
The resulting band structure and Fermi surface are shown in Fig.~\ref{fig:normal}. It can be seen by comparing with the KFe$_2$Se$_2$ bands in Ref.~\onlinecite{Liu:2012} that the modification of the hopping parameters shifts down the hole band near the $\Gamma$ point and leads to a Fermi surface with only electron sheets. The Fermi surface shown in Fig.~\ref{fig:normal}(b,c) corresponds to an electron
doping of K$_x$Fe$_{2-y}$Se$_2$ with $x=0.3$, and $y=0$ (0.15 electrons per Fe) which may be contrasted to the undoped compound with $x=1.0$, and $y=0$ (0.5 electrons per Fe). The Fermi surface calculated in Fig.~\ref{fig:normal}(c) contains a small electron pocket around the Z point in addition to the Fermi pockets around the M point. The LDOS calculated at $k_z=0$ in Fig.\ref{fig:normal}(d) contains mainly $d_{xz}$, $d_{yz}$ and $d_{xy}$ orbitals near the fermi energy whereas the contribution from the $d_{x^2-y^2}$ and $d_z^2$ orbitals near the Fermi energy are negligible.

The second term of Eq.(\ref{eq:H}) describes the intra-site Coulomb interaction
\begin{align}
 \label{eq:Hint}
 H_{int}&=U\sum_{\mathbf{i},\mu}n_{\mathbf{i}\mu\uparrow}n_{\mathbf{i}\mu\downarrow}+(U'-\frac{J}{2})\sum_{\mathbf{i},\mu<\nu,\sigma\sigma'}n_{\mathbf{i}\mu\sigma}n_{\mathbf{i}\nu\sigma'}\\\nonumber
&\quad-2J\sum_{\mathbf{i},\mu<\nu}\vec{S}_{\mathbf{i}\mu}\cdot\vec{S}_{\mathbf{i}\nu}+J'\sum_{\mathbf{i},\mu<\nu,\sigma}c_{\mathbf{i}\mu\sigma}^{\dagger}c_{\mathbf{i}\mu\bar{\sigma}}^{\dagger}c_{\mathbf{i}\nu\bar{\sigma}}c_{\mathbf{i}\nu\sigma},
\end{align}
which includes the intra-orbital (inter-orbital) interaction $U$ ($U'$), the Hund's rule coupling $J$ and the pair hopping energy $J'$.
We will assume orbital and spin rotational invariance where the relations $U'=U-2J$ and $J'=J$ hold.

The third term in Eq.(\ref{eq:H}) is the superconducting pairing term
\begin{equation}
 H_{SC}=-\sum_{\mathbf{i}\neq \mathbf{j},\mu\nu}[\Delta_{\mathbf{ij}}^{\mu\nu}
 c_{\mathbf{i}\mu\uparrow}^{\dagger}c_{\mathbf{j}\nu\downarrow}^{\dagger}+H.c.],
\end{equation}
with the order parameter
$\Delta_{\mathbf{ij}}^{\mu\nu}=\Gamma_{\mu,\nu}^{\mu,\nu}(\mathbf{i},\mathbf{j})\langle c_{\mathbf{j}\nu\downarrow}c_{\mathbf{i}\mu\uparrow}\rangle$.
Here, $\Gamma_{\mu,\nu}^{\mu,\nu}(\mathbf{i},\mathbf{j})$ is the strength of an effective attraction that is generated from a spin fluctuation exchange mechanism. To this end, we follow the procedure given e.g. by Graser {\it et al.}\cite{Graser:2009} and calculate the singlet pairing vertex
\begin{align}
 \label{eq:RPA}
 \Gamma_{st}^{pq}(k,k',\omega)=[{3\over2}U^s\chi_1^{RPA}(k-k',\omega)U^s+{1\over2}U^s\nonumber\\
 -{1\over2}U^c\chi_0^{RPA}(k-k',\omega) U^c+{1\over2}U^c]_{ps}^{tq}.
\end{align}

Note that we have retained only those pairing vertices of the form $\Gamma_{\mu,\nu}^{\mu,\nu}(\mathbf{i},\mathbf{j})$. This is done only for computational speed since we do not find any qualitative changes in our results by including additional inter-orbital interactions as they are quite small in magnitude. The pairing interaction strengths are fixed by requiring that  $U$ and $J$ generate a superconducting gap magnitude of $\Delta=12$meV for both the $d$-wave and $s$-wave symmetry which agrees reasonably well with gap magnitude observed e.g. in ARPES experiments.\cite{Wang:2011} The obtained pairing symmetry is determined simply by the Fermi surface used as input to Eq.(\ref{eq:RPA}), i.e. the Fermi surface in Fig.\ref{fig:normal}(b) [Fig.\ref{fig:normal}(c)] generates $d$ [$s$] pairing, respectively.  In Table~\ref{table:gapvals} we show the self-consistent gap magnitudes $\Delta_{\mathbf{ij}}^{\mu\nu}$ on each orbital. It can be seen that the gap primarily resides on the $d_{xy}$ orbital which we attribute to the intra orbital pairing interaction of the $d_{xy}$ orbital $(\Gamma_{xy,xy}^{xy,xy}(\bold{r},\bold{r'}))$ being at least twice that of the $d_{xz}$/$d_{yz}$ orbital $(\Gamma_{xz,xz}^{xz,xz}(\bold{r},\bold{r'}))$.

\begin{table}[b]
\begin{tabular}{|l| l| l| l| l| l| l|}
  \hline
     Gap & Bond ($\Delta x, \Delta y$) & $d_{xz}$ & $d_{yz}$ & $d_{x^2-y^2}$ & $d_{xy}$ & $d_{z^2}$ \\
  \hline
   $d$-wave & (1,0), (-1,0) & 0.62 & 0.69 & 0.03 & 7.19 & 0 \\
   $d$-wave & (0,1), (0,-1) & -0.69 & -0.62 & -0.03 & -7.19 & 0 \\
   $s$-wave & (1,1), (-1,-1) & 0.68 & 0.68 & 0.02 & 8.68 & 0 \\
   $s$-wave & (-1,1), (1,-1) & 0.68 & 0.68 & 0.02 & 8.68 & 0 \\
  \hline
\end{tabular}
\caption{Real space self-consistent gap values $\Delta_{\mathbf{ij}}^{\mu\nu}$ on each orbital for
nearest neighbor and next nearest neighbor pairings. Units are in meV.}
\label{table:gapvals}
\end{table}

We stress that the present calculation is 2D, and for the results discussed below we use a tight-binding band with the Fermi surface shown in Fig.~\ref{fig:normal}(b). It is only for generation of the pairing couplings $\Gamma$ that we utilise the fact that $d$- or $s$-wave pairing is preferred within the RPA spin-fluctuation approach by Fermi surfaces corresponding to Fig.~\ref{fig:normal}(b) and Fig.~\ref{fig:normal}(c), respectively. This is simply a way to simulate an $s$-wave state in the presence of the Fermi surface of Fig.~\ref{fig:normal}(b). Such a superconducting phase could be stabilised by additional pair scattering processes than the ones allowed by Fig.~\ref{fig:normal}(b) caused by e.g. grazing hole pockets near the $\Gamma$ point, or important inter-band processes between hole and electron processes at $k_z=\pi$.\cite{Hirschfeld:2011}

The last term in the Hamiltonian (\ref{eq:H}) incorporates the iron vacancies.
We model the vacancies as impurities with sufficiently large onsite potentials that remove any electronic states from their location
\begin{equation}
 H_{vacancy}=V_{vacancy}\sum_{\mathbf{i^*}\mu\sigma}c_{\mathbf{i^*}\mu\sigma}^{\dagger}c_{\mathbf{i^*}\mu\sigma},
\end{equation}
which adds the potential $V_{vacancy}$ at the vacancy sites $\mathbf{i^*}$.

After a standard mean-field decoupling of the onsite interaction term \eqref{eq:Hint} we arrive at the following multiband Bogoliubov de-Gennes equations\cite{Gastiasoro:2013}
\begin{align}
\sum_{\mathbf{j}\nu}
\begin{pmatrix}
H_{\mathbf{i}\mu \mathbf{j}\nu \sigma} & \Delta_{\mathbf{i}\mu \mathbf{j}\nu}\\
\Delta_{\mathbf{i}\mu \mathbf{j}\nu}^{*} & -H_{\mathbf{i}\mu \mathbf{j}\nu \bar{\sigma}}^{*}
\end{pmatrix}
\begin{pmatrix}
 u_{\mathbf{j}\nu}^{n} \\ v_{\mathbf{j}\nu}^{n}
\end{pmatrix}=E_{n}
\begin{pmatrix}
 u_{\mathbf{i}\mu}^{n} \\ v_{\mathbf{i}\mu}^{n}
\end{pmatrix},
\end{align}
where
\begin{align}
 H_{\mathbf{i}\mu \mathbf{j}\nu \sigma}&=t_{\mathbf{ij}}^{\mu\nu}+\delta_{\mathbf{ij}}\delta_{\mu\nu}[-\mu_{0}+\delta_{\mathbf{ii^*}}V_{vacancy}+U \langle n_{\mathbf{i}\mu\bar{\sigma}}\rangle\\\nonumber
&\quad+\sum_{\mu' \neq \mu}(U'\langle n_{\mathbf{i}\mu' \bar{\sigma}}\rangle+(U'-J)\langle n_{\mathbf{i}\mu' \sigma}\rangle)].
\end{align}
The local densities and the SC order parameters are obtained self-consistently through iteration of
\begin{align}
  \langle n_{\mathbf{i}\mu\uparrow} \rangle&=\sum_{n}|u_{\mathbf{i}\mu}^{n}|^{2}f(E_{n}),\\\nonumber
  \langle n_{\mathbf{i}\mu\downarrow} \rangle&=\sum_{n}|v_{\mathbf{i}\mu}^{n}|^{2}(1-f(E_{n})),\\
   \Delta_{\mathbf{ij}}^{\mu\nu}&=\Gamma_{\mu,\nu}^{\mu,\nu}(\mathbf{i},\mathbf{j})\sum_{n}u_{\mathbf{i}\mu}^{n}v_{\mathbf{j}\nu}^{n*}f(E_{n}).
\end{align}
In the following section, we discuss the results of the above procedure for the magnetic, and superconducting properties
applicable to iron chalcogenide superconductor K$_x$Fe$_{2-y}$Se$_2$.

\section{Results}

\subsection{Normal state}

The band structure properties of the iron based superconductors generally include contributions
 from all five $d$-orbitals near the Fermi surface. A simple electron count and ARPES results reveals that the 122 iron selenides
 are strongly electron doped.
 We impose a $15\%$ electron doping on the metallic region by adjusting the chemical potential to $\mu=-0.23$eV. The
 evaluated Fermi surface shape for $k_z=0$ and $k_z=\pi$ are shown in Fig.~\ref{fig:normal}(b) and \ref{fig:normal}(c), respectively. They agree well with ARPES
 observations which find only electron pockets at the M points for $k_z=0$ and a small additional electron pocket around the Z point.\cite{Wang:2011, Zhang:2011,Qian:2011} In our calculations for the normal metallic state, the onsite Coulomb term has been fixed by $U=0.6$eV and $J=0.25U$.

 In Fig.~\ref{fig:normal}(d) we show the orbitally resolved LDOS. It can be seen that the primary contributions to
 the density of states near the Fermi level are from the $d_{xy}$ and $d_{xz}/d_{yz}$ orbitals. Though the orbital content can
 depend on the doping level as well as the presence of iron vacancies, observations on similar alkali doped iron selenide systems identify
 the above t$_{2g}$ orbitals to be the dominant contributors of the Fermi surface at $k_z=0$.\cite{Fei:2012}

\subsection{Block antiferromagnetic state}

\begin{figure}[t]
\begin{minipage}{.98\columnwidth}
\includegraphics[clip=true,width=0.99\columnwidth]{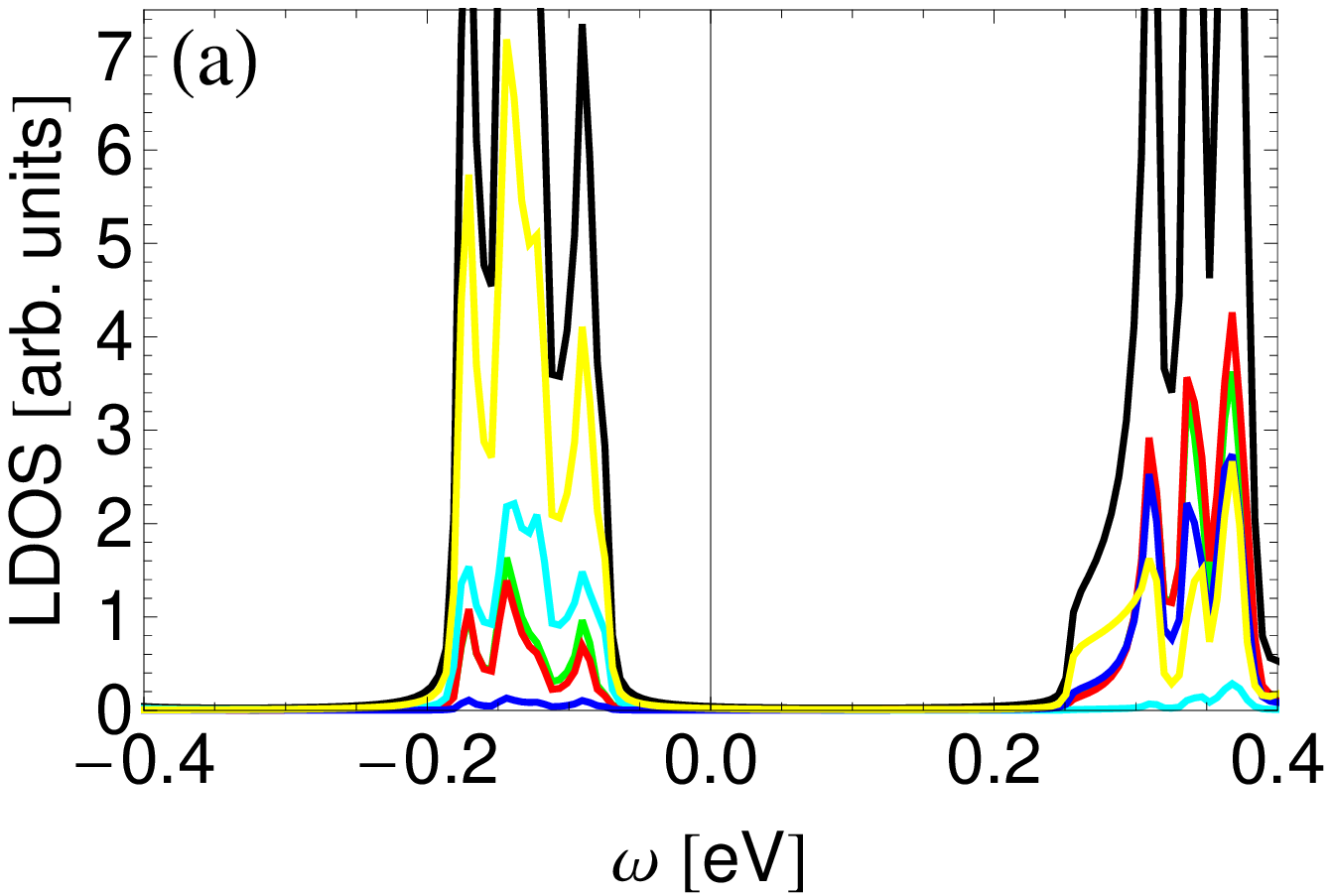}
\end{minipage}\\
\begin{minipage}{.49\columnwidth}
\includegraphics[clip=true,width=0.99\columnwidth]{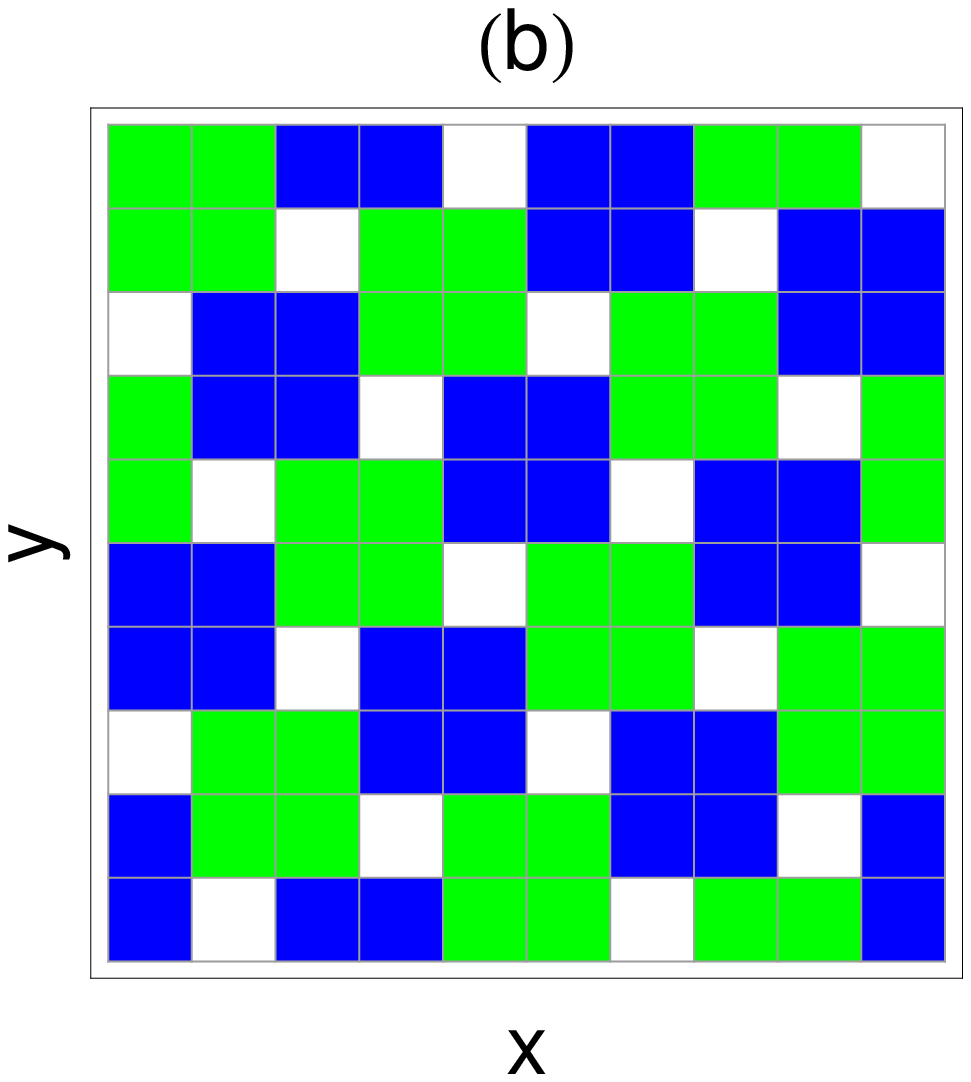}
\end{minipage}
\begin{minipage}{.49\columnwidth}
\includegraphics[clip=true,width=0.99\columnwidth]{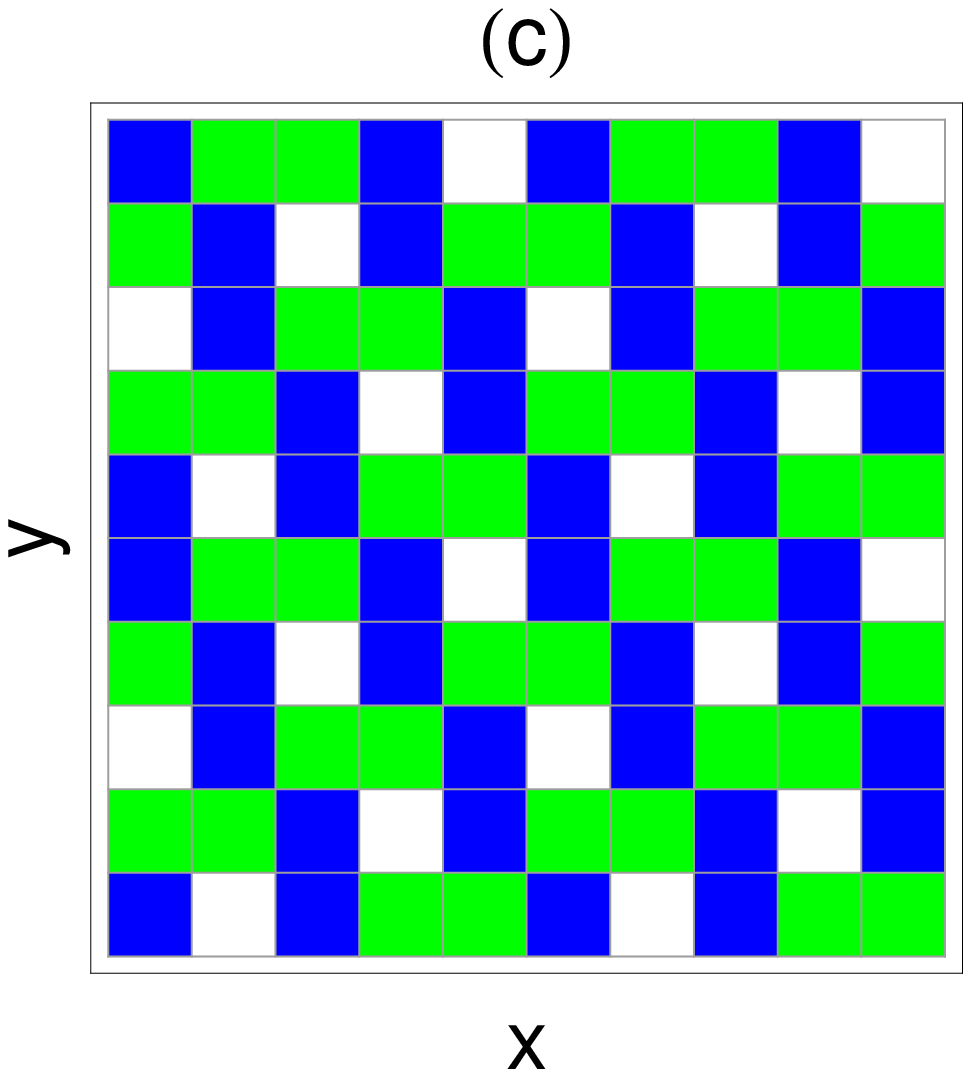}
\end{minipage}
\caption{(Color online) (a) Orbital resolved LDOS of the BAFM region with a normal state band corresponding to Fig.~\ref{fig:normal}. (b) Real-space map of the magnetic moment in the BAFM state. White: vacancy sites, green: 3.63$\mu_B$/Fe, blue: -3.63$\mu_B$/Fe.
(c) Same as (b) but for orbital order $n_{xz}$-$n_{yz}$. Blue: -0.011, green: 0.011.}
\label{fig:bafm}
\end{figure}

The presence of an ordered arrangement of iron vacancies with a $\sqrt{5}\times\sqrt{5}$ arrangement in the lattice structure
of K$_x$Fe$_{2-y}$Se$_2$ stabilises a block antiferromagnetic state (BAFM). STM measurements, for example, find that these regions are insulating.\cite{Li(a):2012} Additionally, experiments find an electron occupation of 6e$^{-}/$Fe.\cite{Yan:2012,Texier:2012}

 In our model calculations, the chemical potential for the BAFM state is the same as in the normal metal discussed above. An electron density of  6 e$^{-}$/Fe is obtained by a Coulomb interaction strength of $U=1.35$eV and $J = 0.25 U$ in the BAFM state. The stoichiometry for the BAFM with $\sqrt{5}\times\sqrt{5}$ can be considered to be K$_{0.8}$Fe$_{1.6}$Se$_2$ from an electron and vacancy concentration count.
 It has been previously argued by Luo {\it et al.}~\cite{Luo:2012} that a correct description of the magnetic state requires a larger Coulomb
 interaction term compared to what is typically used for pnictides. In addition, we note that in the phase separated region the normal metal is likely
 to have a smaller effective on-site Coulomb repulsion as otherwise the normal region would undergo a SDW transition to a $(\pi,\pi)$
 magnetic state and no such magnetism has been observed.

 Using the self-consistency approach discussed in section 2 we find a stable BAFM state with a magnetic moment of 3.63 $\mu_B$/Fe.
 The magnetic moment obtained is shown in a real-space map in Fig.~\ref{fig:bafm}(b). Calculation of the associated LDOS yields an insulating gap
  of around 0.35 eV as shown in Fig.~\ref{fig:bafm}(a). This gap agrees reasonably well with the gap magnitude obtained in STM measurements of the BAFM regions.\cite{Li(a):2012} Another result that comes out naturally from this calculation is the presence of orbital order defined as $n_{xz}-n_{yz}$ as shown in Fig.~\ref{fig:bafm}(c). The origin of the orbital order is the orbitally selective suppression of hopping of an electron in a $d_{xz}$ ($d_{yz}$) orbital in the horizontal (vertical) direction next to a vacancy location.\cite{Lv:2011}

\subsection{Superconducting state}

 \begin{figure}[t]
\begin{minipage}{.49\columnwidth}
\includegraphics[clip=true,width=0.99\columnwidth]{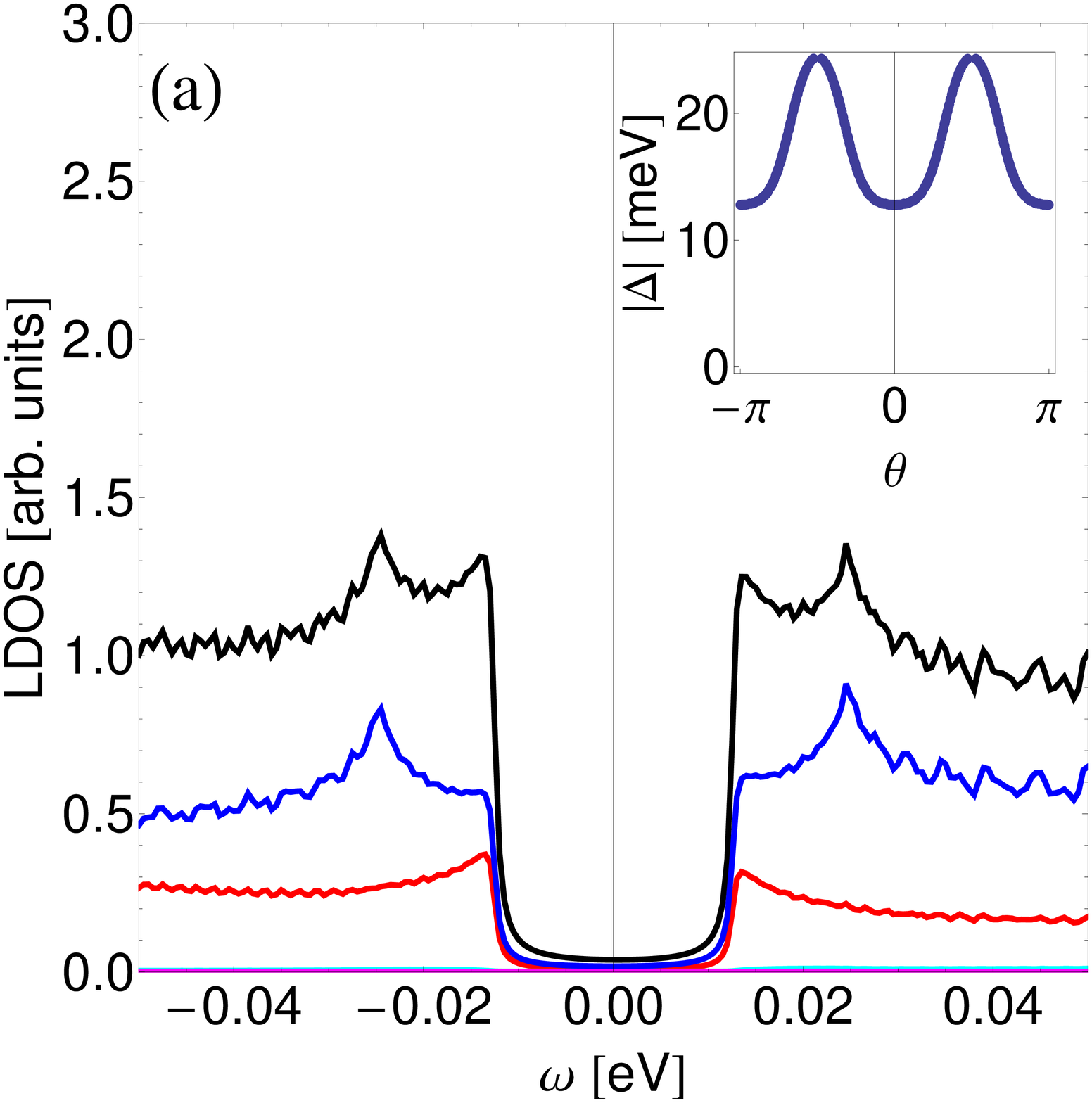}
\end{minipage}
\begin{minipage}{.49\columnwidth}
\includegraphics[clip=true,width=0.99\columnwidth]{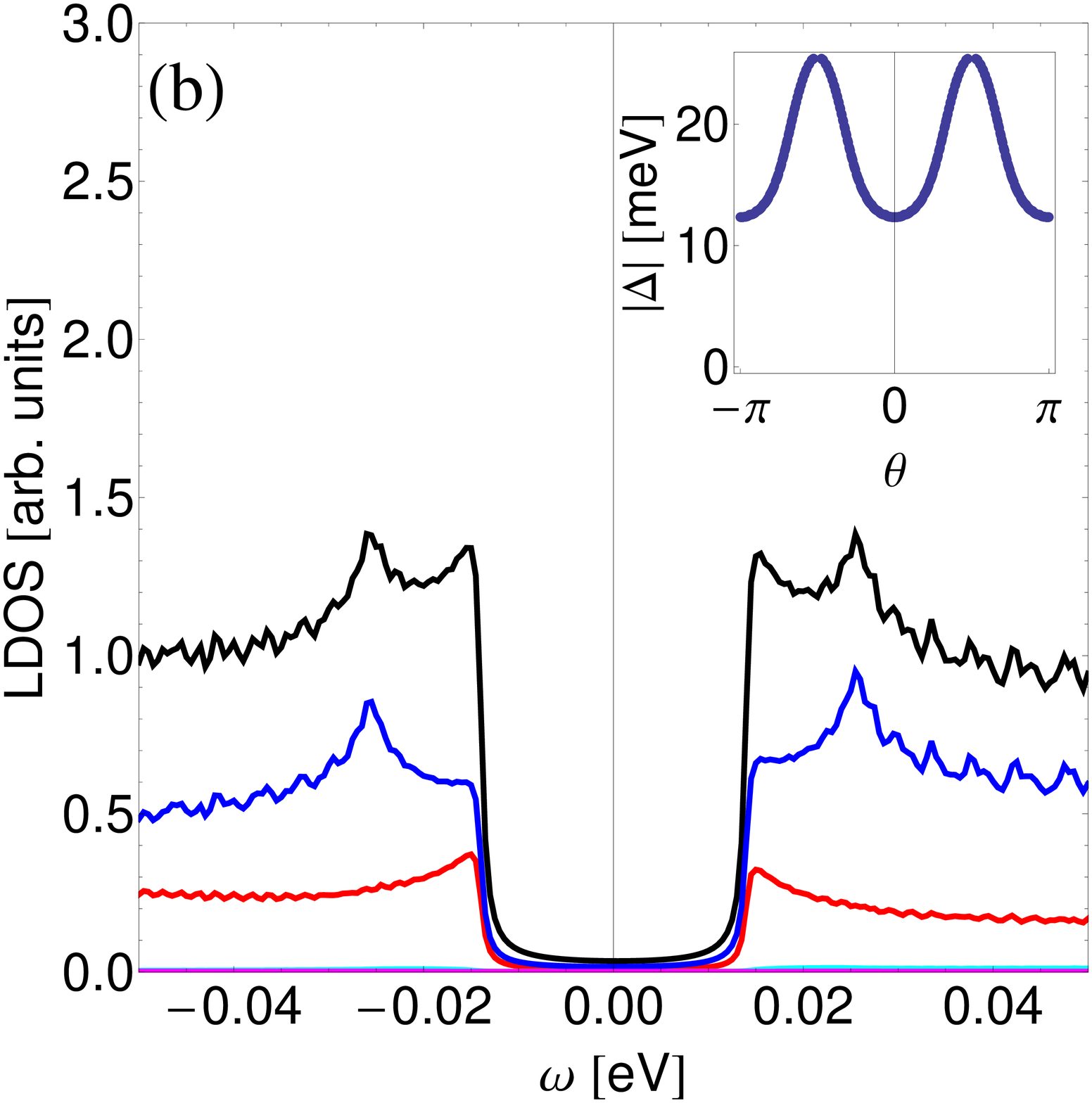}
\end{minipage}
\caption{(Color online) (a) LDOS in $d$-wave superconducting state. Black: total LDOS, blue: $d_{xy}$, red: $d_{xz}/d_{yz}$. The inset shows the gap anisotropy around the Fermi surface centered at $(0,\pi)$. The $x$-axis is angle measured with respect to the $k_x$ direction.
(b) Same as (a) but for the $s$- wave symmetry.}
\label{fig:sc}
\end{figure}
As described in the model section, superconducting order is stabilised by including the pairing interactions generated within a standard weak coupling RPA method.
Before turning to the discussion of the LDOS near boundary regions between superconducting and BAFM regions, it is instructive to briefly display the properties of the homogeneous superconductor. In Fig.~\ref{fig:sc} we show the orbitally resolved LDOS for the superconducting state with $d$-wave and $s$-wave pairing symmetry, respectively. Due to the same gap amplitude in the two cases, and the fact that the LDOS is not phase sensitive, the final LDOS of the homogeneous phase is very similar in the two states as expected. The calculated LDOS are typical for a nodeless superconducting state and display an apparent "two-gap" behavior similar to what has been observed in recent experiments.\cite{Li(a):2012,Li(b):2012} The two sets of coherence peaks close to $\sim 12$meV and $\sim 25$meV in Fig.~\ref{fig:sc} results from a significant gap anisotropy on the electron pockets as seen from the insets in Figs.~\ref{fig:sc}.
The superconducting pairing interaction has been chosen to result in minimum gap amplitudes of about 12meV which is of the order of experimentally observed amplitudes that have been observed to be around 10meV.

\subsection{Interface between BAFM and superconductor}

\subsubsection{BAFM/SC interface}

\begin{figure}[]
\begin{minipage}{.99\columnwidth}
\includegraphics[clip=true,width=0.99\columnwidth]{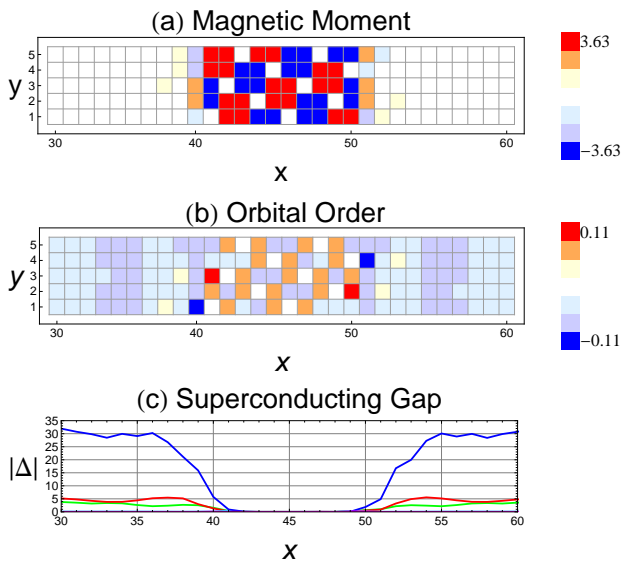}
\end{minipage}
\caption{(Color online) Spatial dependence of the order parameters near an interface between a BAFM and $d$-wave superconductor. The system size is $80\times10$ with a BAFM of $10\times10$ sites placed between site $x=40$ and $x=50$. (a) Map of the magnetic moment ($\mu_B/$Fe) shown between sites $x=30$ to $x=60$ and $y=1$ to $y=5$.
(b) Same as (a) but for the orbital order ($n_{xz}-n_{yz}$).
(c) Orbital resolved averaged superconducting gap magnitude shown between $x=30$, $y=10$ to $x=60$, $y=10$. The color scheme is the same as in Fig.~\ref{fig:normal}.}
\label{fig:interface1}
\end{figure}

\begin{figure}
\begin{minipage}{.49\columnwidth}
\includegraphics[clip=true,width=0.99\columnwidth]{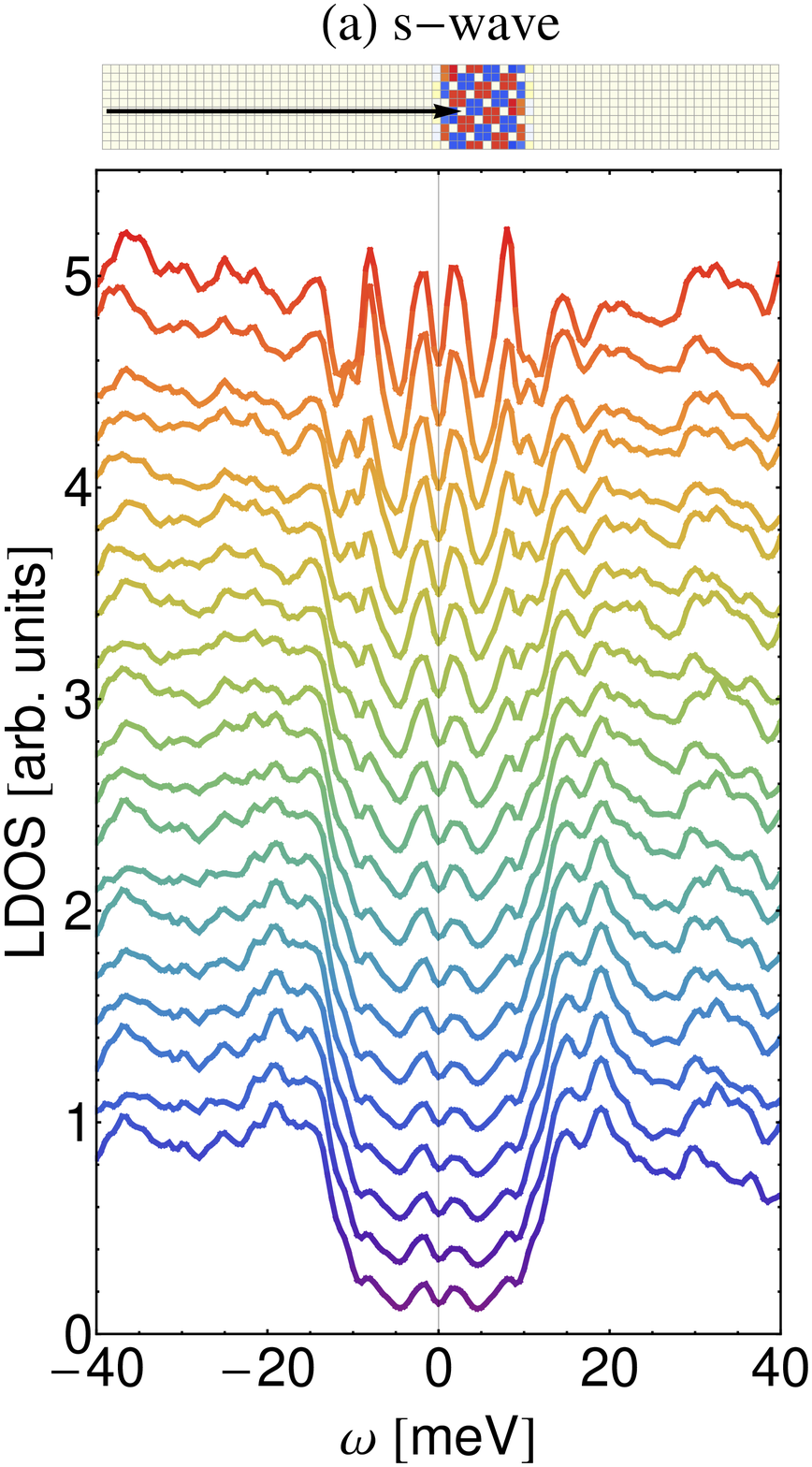}
\end{minipage}
\begin{minipage}{.49\columnwidth}
\includegraphics[clip=true,width=0.99\columnwidth]{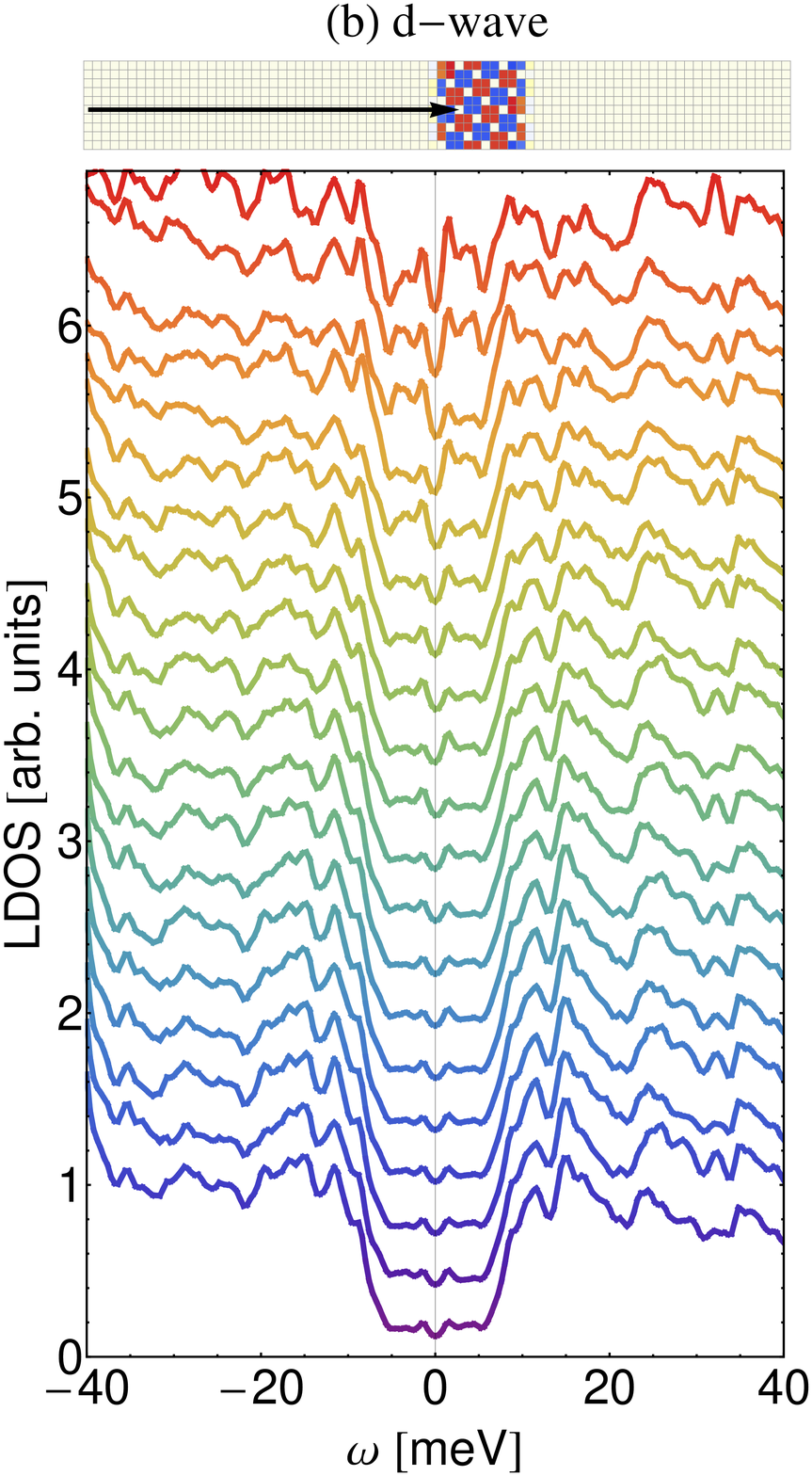}
\end{minipage}
\caption{(Color online) (a) Evolution of the LDOS upon approaching a BAFM interface from $x=1$ to $x=42$ for $s$-wave gap symmetry. (b) Same as (a) but for $d$-wave symmetry of the superconducting order parameter.}
\label{fig:interface}
\end{figure}

As mentioned in the introduction, the alkali doped iron selenide materials A$_x$Fe$_{2-y}$Se$_2$ are phase separated, and interfaces between BAFM and superconducting regions are abundant and an important situation to understand in detail. For such an interface, it can be seen from Fig.~\ref{fig:interface1}(a) that the magnetism penetrates only
a few lattice spacings into the superconducting region due to the significant amplitude of the magnetisation in the BAFM regions. Since the BAFM is insulating, the superconducting order parameter also does not exhibit any significant leaking into the BAFM regions as seen from Fig.~\ref{fig:interface1}(c). This short-range proximity effect is qualitatively consistent with the existence of filamentary superconducting channels near these magnetic regions.\cite{Ding:2012,Ricci:2011,Yuan:2012} The orbital
order near the interface is shown in Fig.~\ref{fig:interface1}(b). As seen, at particular sites near the interface, the orbital
order undergoes significant enhancement by a factor of about ten compared to the size of the orbital order in the bulk BAFM. This takes place adjacent to the vacancy sites bordering the superconductor causing significant local LDOS modulations at these sites. Note that the absolute change in the orbital order near the interface is still weak with $n_{xz}-n_{yz}\sim 0.11$.

What are the low-energy states near the BAFM and superconductor interfaces? In Fig.~\ref{fig:interface}  we show the evolution of the LDOS upon approaching the interface for both the $s$- and $d$-wave cases. As seen in both cases, prevalent low-energy in-gap bound states clearly exist, and there is no qualitative difference in the LDOS caused by the pairing symmetry of the bulk superconductor. This is contrary to the case of a single nonmagnetic impurity where Zhu {\it et al.}\cite{Zhu:2011} showed that the $d$-wave state leads to in-gap bound states as opposed to the nnn $s$-wave state (see however Ref.~\onlinecite{Hirschfeld:2011, Beaird:2012} for a more  general discussion). The magnetisation of the BAFM is crucial for the generation of in-gap states in both Fig.~\ref{fig:interface}(a) and  Fig.~\ref{fig:interface}(b) since it breaks time-reversal symmetry and generates bound states for both $s$- and $d$-wave superconductors. Certainly, in a picture where one thinks of the interface as a row of magnetic and non-magnetic impurities the resulting interference of the multiple generated bound states should generate an in-gap band of states leading to LDOS results similar to those shown in Fig.~\ref{fig:interface}.\cite{morr,andersen3}

\subsubsection{Scattering off a (110) wall}

\begin{figure}
\begin{minipage}{.49\columnwidth}
\includegraphics[clip=true,width=0.99\columnwidth]{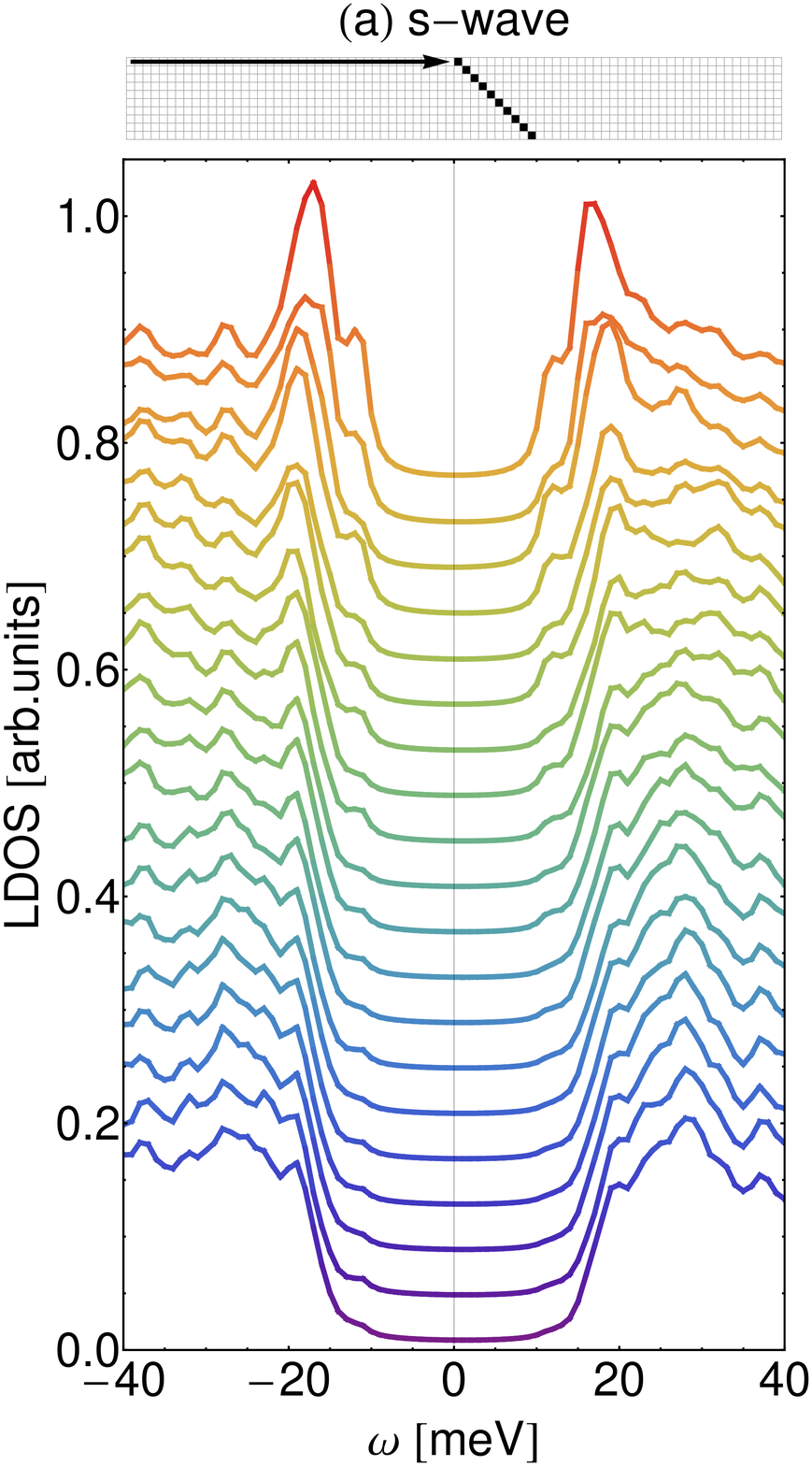}
\end{minipage}
\begin{minipage}{.49\columnwidth}
\includegraphics[clip=true,width=0.99\columnwidth]{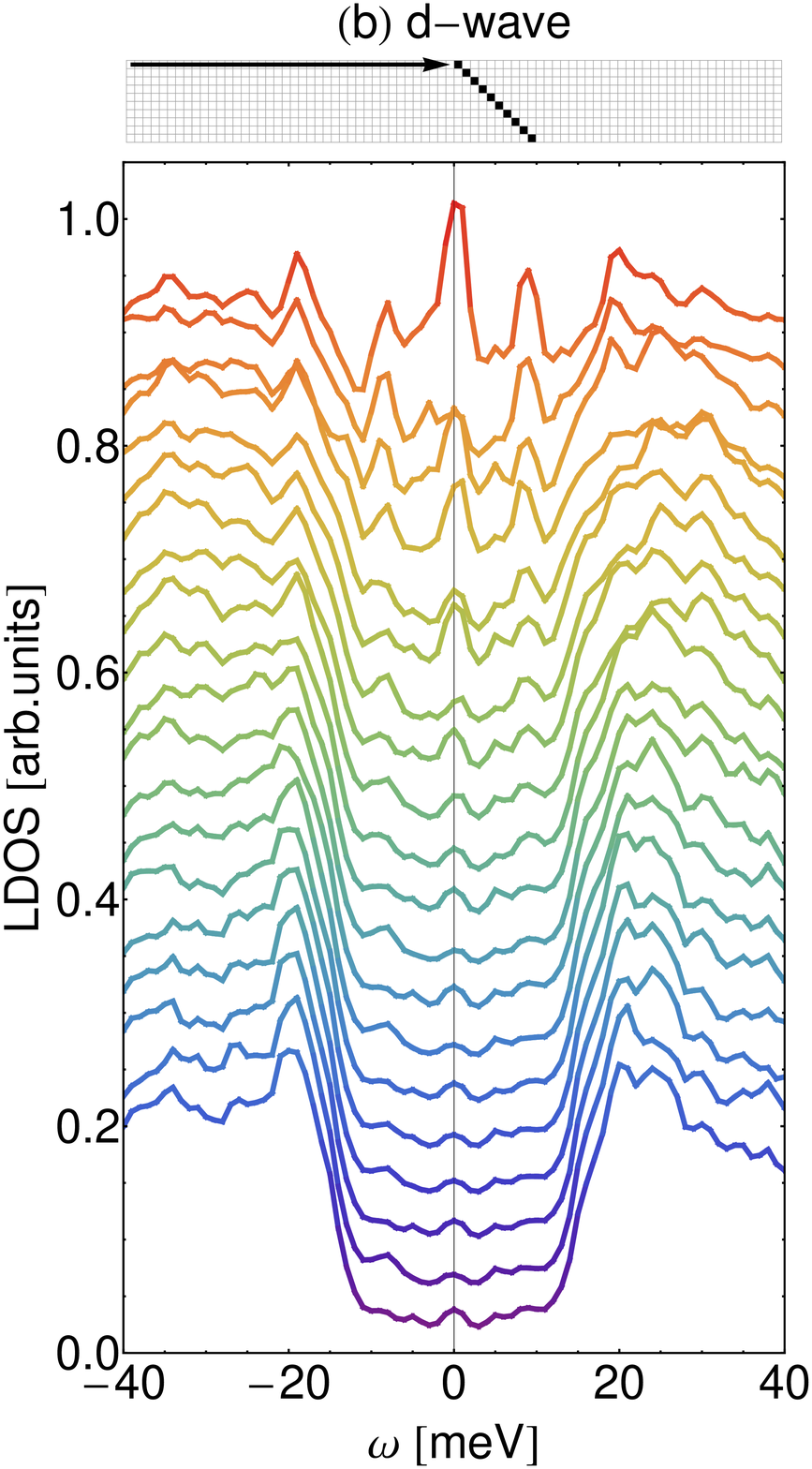}
\end{minipage}
\begin{minipage}{.49\columnwidth}
\includegraphics[clip=true,width=0.99\columnwidth]{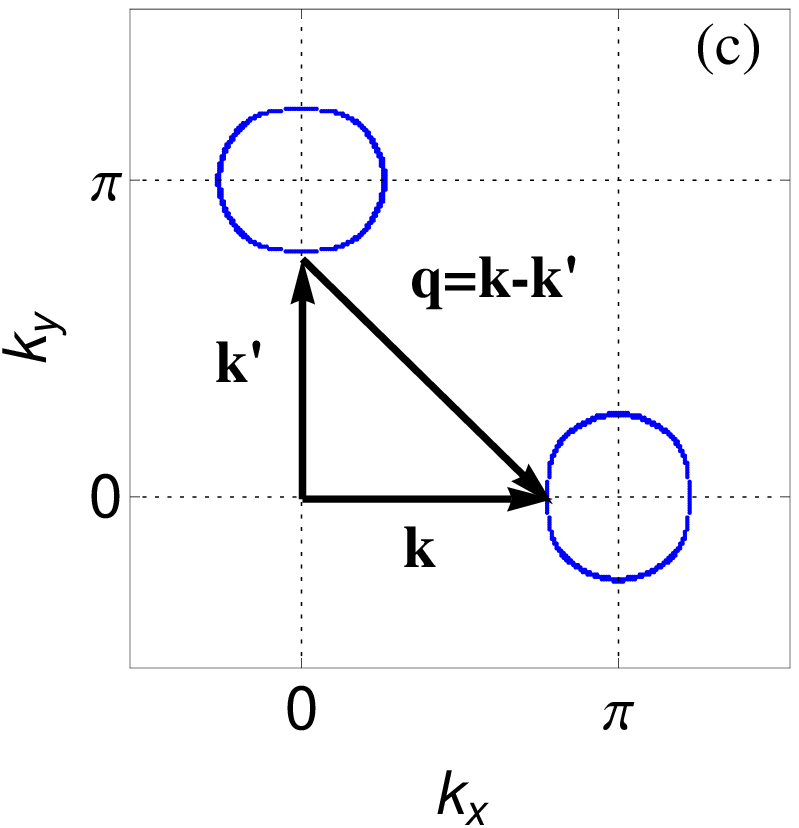}
\end{minipage}
\begin{minipage}{.49\columnwidth}
\includegraphics[clip=true,width=0.99\columnwidth]{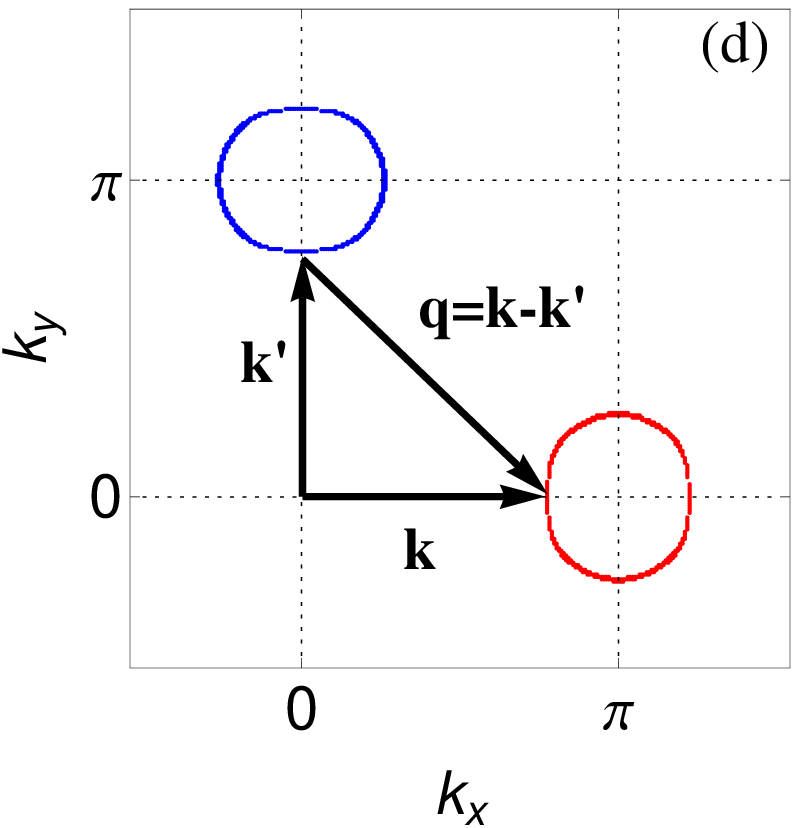}
\end{minipage}
\caption{(Color online) An 80 by 10 system with an interface between an $s$-wave nnn superconductor and a BAFM with vacancies forming a (110) wall.
Here, $U=0.6$, and $J=0.25U$ which leads to a nonmagnetic interface. (a) LDOS plots for $x=1$ to $x=40$ for $s$-wave gap symmetry. (b) Same as (a) but for $d$-wave gap symmetry. (c) Incident wave vector $\mathbf{k}$ at the (110) interface and reflected wave vector $\mathbf{k'}$ for $s$-wave symmetry. (d) Same as in (c) but for $d$-wave symmetry. Blue: $|\Delta|$, Red: -$|\Delta|$. }
\label{fig:diagint}
\end{figure}

In the case of cuprate $d$-wave superconductors it is well-known that a (110) diagonal surface/interface or grain boundary leads to the generation of a zero-energy resonant state due to a sign change of the gap function between the incoming and scattered momentum vector within a semiclassical picture.\cite{hu,tanaka,buchholtz} In the present case, a similar scattering geometry which probes the different gap signs of the electron pockets should provide a qualitatively different low-energy LDOS between the $s$- and $d$-wave superconductors.  As shown in Fig.~\ref{fig:diagint}, a (110) wall placed at 45 degrees to the $x$-axis indeed leads to a bound state solutions for the $d$-wave state unlike the case for the $s$-wave gap symmetry. The origin of this difference is exactly the same as for the standard one-band $d$-wave materials.\cite{hu,tanaka,buchholtz,andersen} Thus, using scanning probes to locate such dislocations in the FeSe planes, local STM measurements near such regions could provide important clues to the pairing symmetry of these materials.

\section{conclusions}

We have provided a microscopic real-space study of the interplay between BAFM and superconducting domains relevant to alkali doped iron selenide materials A$_x$Fe$_{2-y}$Se$_2$ which have been shown to exhibit significant in-plane and out-of-plane phase separation. Our modelling includes all five iron $d$-orbitals and the onsite Coulomb interaction treated at the mean field level. Superconductivity is incorporated through the spin-fluctuation RPA approach. A main result is that low-energy in-gap bound states should exist near interfaces between BAFM and superconducting regions irrespective of whether the pairing symmetry of the superconductor is $d$- or $s$-wave. By contrast, a non-magnetic (110) scattering plane only causes low-energy bound states for the $d$-wave pairing and should provide a means to distinguish between $d$ and $s$-wave pairing symmetry of these materials.

\section{acknowledgements}
B.M.A. and M.N.G. acknowledge support from The Danish Council for Independent Research $|$ Natural Sciences, and the Lundbeckfond fellowship (grant A9318).


\begin{thebibliography}{100}
\bibitem{Hirschfeld:2011} P. J. Hirschfeld, M. M. Korshunov, and I. I. Mazin, Rep. Prog. Phys. {\bf 74}, 124508 (2011).
%
\bibitem{Dagotto:2012} E. Dagotto,  arXiv:1210.6501 (2012).
%
\bibitem{Guo:2010} J. Guo, S. Jin, G. Wang, S. Wang, K. Zhu, T. Zhou, M. He, and X. Chen, Phys. Rev. B {\bf 82}, 180520 (2010).
%
\bibitem{Zavalij:2011} P. Zavalij, W. Bao, X. F. Wang, J. J. Ying, X. H.  Chen, D. M.  Wang, J. B.  He, X. Q. Wang, G. F. Chen, P.-Y. Hsieh, Q.  Huang, and M. A. Green, Phys. Rev. B  {\bf 83}, 132509 (2011).
%
\bibitem{Bao:2011} W. Bao, Q. Huang, G. F. Chen, M. A. Green, D. M. Wang, J. B. He, X. Q. Wang, and Y. Qiu,
Chin. Phys. Lett. {\bf 28}, 086104 (2011).
%
\bibitem{Pomjakushin:2011} V. Yu Pomjakushin, E. V. Pomjakushina, A Krzton-Maziopa, K Conder, and Z. Shermadini, J. Phys. Condens. Matter, {\bf 23}, 156003 (2011).
%
\bibitem{Ye:2011} F. Ye, S. Chi, Wei Bao, X. F. Wang, J. J. Ying, X. H. Chen, H. D. Wang, C. H. Dong, and M. Fang, Phys. Rev. Lett. {\bf 107}, 137003 (2011).
%
\bibitem{Mazin:2011} I., Mazin Physics {\bf 4}, 26 (2011).
%
\bibitem{Chen:2011} F. Chen, M. Xu, Q. Q. Ge, Y. Zhang, Z. R. Ye, L. X. Yang, Juan Jiang, B. P. Xie, R. C. Che, M. Zhang, A. F. Wang, X. H. Chen,  D. W. Shen, J. P. Hu, and D. L. Feng, Phys. Rev. X {\bf1}, 021020 (2011).
%
\bibitem{Lazarevic:2012} N. Lazarevi\ifmmode \acute{c}\else \'{c}\fi{}, M. Abeykoon, P. W. Stephens, Hechang Lei, E. S. Bozin, C. Petrovic, and Z. V. Popovi\ifmmode \acute{c}\else \'{c}\fi{}, Phys. Rev. B {\bf86}, 054503 (2012).
%
\bibitem{Yuan:2012} R. H. Yuan, T. Dong, Y. J. Song, P. Zheng, G. F. Chen, J. P. Hu, J. Q. Li, and N. L. Wang, Sci. Rep. {\bf2}, 221 (2012).
%
\bibitem{Ding:2012} X. Ding, D. Fang, Z. Wang, H. Yang, J. Liu, Q. Deng, G. Ma, C. Meng, Y. Hu, and H.- H. Wen, arXiv:1301.2668.
%
\bibitem{Li(a):2012} W. Li, H. Ding, P. Deng, K. Chang, C. L. Song, K. He, L. L. Wang, X. C. Ma, J. P.  Hu, X. Chen, and Q. K. Xue, Nature Phys. {\bf 8} 126 (2012).
%
\bibitem{Li(b):2012} W. Li, H. Ding, Z. Li, P. Deng, K. Chang, K. He, S. Ji, L. Wang, X. Ma, J.-P. Hu, X. Chen, and Q.-K. Xue, Phys. Rev. Lett. {\bf 109}, 057003 (2012).
%
\bibitem{Yan:2012} Y. J. Yan, M. Zhang, A. F. Wang, J. J. Ying, Z. Y. Li, W. Qin, X. G. Luo, J. Q. Li, J. Hu, and X. H. Chen, Sci. Rep. {\bf2}, 212 (2012).
%
\bibitem{Landsgesell:2012} S. Landsgesell, D. Abou-Ras, T. Wolf, D. Alber, and K.  Prokes, Phys. Rev. B {\bf 86}, 224502 (2012).
%
\bibitem{Charnukha:2012} A. Charnukha, A. Cvitkovic, T. Prokscha, D. Pr\"opper, N. Ocelic, A. Suter, Z. Salman, E. Morenzoni, J. Deisenhofer, V. Tsurkan, A. Loidl, B. Keimer, A. V. Boris, Phys. Rev. Lett. {\bf 109}, 017003 (2012).
%
\bibitem{Ricci:2011} A. Ricci, N. Poccia, G. Campi, B. Joseph, G. Arrighetti, L. Barba, M. Reynolds, M. Burghammer, H. Takeya, Y.  Mizuguchi, Y. Takano, M. Colapietro, N. L.  Saini, and A. Bianconi, Phys. Rev. B. {\bf 84} 060511 (2011).
%
\bibitem{Liu:2012} Y. Liu, Q. Xing, K. W. Dennis, R. W. McCallum, and T. A. Lograsso, Phys. Rev. B {\bf86}, 144507 (2012).

%
\bibitem{Texier:2012} Y. Texier, J. Deisenhofer, V. Tsurkan, A. Loidl, D. S.  Inosov, G. Friemel, J.  Bobroff, Phys. Rev. Lett. {\bf108}, 237002 (2012).
%
\bibitem{Wang:2011}  Y. Zhang, X. Yang, M. Xu, Z. R. Ye, F. Chen, C. He, H. C. Xu, J. Jiang, B. P. Xie, J. J. Ying,
 X. F. Wang, X. H. Chen, J. P. Hu, M. Matsunami, S. Kimura, and D. L. Feng, Nature Mat. {\bf 10}, 273 (2011).
%
\bibitem{Zhang:2011}  Y. Zhang, L. X. Yang, M. Xu, Z. R. Ye, F. Chen, C. He, H. C. Xu, J. Jiang, B. P. Xie, J. J. Ying, X. F. Wang, X. H. Chen, J. P. Hu, M. Matsunami, S. Kimura, and D. L. Feng, Nature Mat. {\bf 10}, 273 (2011).
%
\bibitem{Qian:2011} T. Qian, X.-P. Wang, W.-C. Jin, P. Zhang, P. Richard, G. Xu, X. Dai, Z. Fang, J.-G. Guo, X.-L. Chen, H. Ding, Phys. Rev. Lett. {\bf106}, 187001 (2011).
%
\bibitem{Mazin:2008} I. I. Mazin, D. J. Singh, M. D. Johannes, and M. H. Du, Phys. Rev. Lett.  {\bf 101}, 057003 (2008).
%
\bibitem{Maier:2011} T. A. Maier, S. Graser, P. J. Hirschfeld, and D. J. Scalapino, Phys. Rev. B  {\bf 83}, 100515 (2011).
%
\bibitem{WangF:2011} F. A. Wang, F. Yang, M. Gao, Z.-Y. Lu, T. Xiang, and D.- H. Lee, Europhys. Lett. {\bf 93}, 57003 (2011).
%
\bibitem{Fang:2011} C. Fang, Y.-L. Wu, R. Thomale, B. A. Bernevig, and J. P. Hu, Phys. Rev. X  {\bf 1}, 011009 (2011).
%
\bibitem{Mazinb:2011} I. I. Mazin Phys. Rev. B {\bf 84}, 024529 (2011).
%
\bibitem{Zeng:2011} B. Zeng, B. Shen, G. F. Chen, J. B. He, D. M. Wang, C. H. Li, and H. H. Wen, Phys. Rev. B {\bf 83}, 144511 (2011).
%
\bibitem{Xu:2012} M. Xu, Q.Q. Ge, R. Peng, Z. R. Ye, J. Jiang, F. Chen, X. P. Shen, B. P. Xie, Y. Zhang, A. F. Wang, X. F. Wang, X. H. Chen, and D. L. Feng, Phys. Rev. B. {\bf 85}, 220504 (2012).
%
\bibitem{Yong:2012} D.- Y. Liu, Y.- M. Quan, Z., Zeng and L.- J. Zou, Physica B  {\bf 407}, 1139 (2012).
%
\bibitem{Graser:2009} S. Graser, T. A. Maier, P. J. Hirschfeld, and D. J. Scalapino, New J. Phys. {\bf 11}, 025016 (2009).
%
\bibitem{Yu:2011} W. Yu, L. Ma, J. B. He, D. M. Wang, T.-L. Xia, G. F. Chen, and W. Bao, Phys. Rev. Lett. {\bf 106}, 197001 (2011).
%
\bibitem{Zhu:2011} J.-X. Zhu, R. Yu, A. V. Balatsky, and Q. Si, Phys. Rev. Lett.  {\bf 107}, 167002 2011.
%
\bibitem{Zhu:2012} J.-X. Zhu, and A. R. Bishop, Europhys. lett. {\bf 100}, 37004 (2012).
%
\bibitem{Lib:2012} Y. K. Li, C. Y. Shen, H. J. Guo, C. Lv, X. J. Yang, L. Zhang, Y. K. Luo, G. H. Cao, and Z. A. Xu J. Phys.: Condens. Matter {\bf 24}, 232202 (2012).
%
\bibitem{Gastiasoro:2013} M. N. Gastiasoro and B. M. Andersen, J. Supercond. Novel. Magn. (2013).
%
\bibitem{Fei:2012} C. Fei, G. QingQin, X. Min, Z. Yan, S. XiaoPing, LI Wei, M. Masaharu, K. Shin-ichi, H. J. Hu, and F. D. Lai, Chin. Sci. Bull. {\bf57} 30 (2012).
%
\bibitem{Cao:2011} C. Cao, and J. Dai, Phys. Rev. Lett. {\bf 107}, 056401 (2011).
%
\bibitem{Lic:2012} W. Li, H. Ding, P. Deng, K. Chang, C. Song, K. He, L. Wang, X. Ma, J.-P. Hu, X. Chen, and Q.-K. Xue,
Nat. Phys. {\bf 8}, 126 (2012).
%
\bibitem{Luo:2012} Q. Luo, A. Nicholson, J. Riera, D.- X. Yao, A. Moreo, and E. Dagotto, Phys. Rev. B {\bf 84}, 140506 (R) (2011).
%
\bibitem{Lv:2011} W. Lv, W.- C. Lee, and P. Phillips, Phys. Rev. B {\bf 84}, 155107 (2011).
%
\bibitem{Beaird:2012} R. Beaird, I. Vekhter, and J.- X. Zhu, Phys. Rev. B. {\bf 86}, 140507 (R) (2012).
%
\bibitem{morr} D. K. Morr and N. A. Stavropoulos, Phys. Rev. B {\bf 66}, 140508 (2002).
%
\bibitem{andersen3} B. M. Andersen and P. Hedeg\aa rd, Phys. Rev. B {\bf 67}, 172505 (2003).
%
\bibitem{hu} C. R. Hu, Phys. Rev. Lett. {\bf 72}, 1526 (1994).
%
\bibitem{tanaka} Y. Tanaka and S. Kashiwaya, Phys. Rev. Lett. {\bf 74}, 3451 (1995).
%
\bibitem{buchholtz} L. J. Buchholtz, M. Palumbo, D. Rainer, and J. A. Sauls, J. Low Temp. Phys. {\bf 101}, 1099 (1995).
%
\bibitem{andersen} B. M. Andersen, I. V. Bobkova, P. J. Hirschfeld, and Yu. S. Barash, Phys. Rev. B {\bf 72}, 184510 (2005).

\end{thebibliography}
\end{document}